\documentclass[11pt]{article}
\usepackage{fa2023}
\usepackage{amsmath}
\usepackage{cite}
\usepackage{url}
\usepackage{graphicx}
\usepackage{color}
\usepackage{siunitx}
\usepackage[utf8]{inputenc}

\usepackage[subtle, charwidths,leading,indent,lists,bibnotes]{savetrees}

\usepackage{flushend}

\usepackage{babel}
\usepackage{subcaption}
\usepackage{tikz}
\usetikzlibrary{positioning}

\title{Speech-dependent Modeling of Own Voice Transfer Characteristics for In-ear Microphones in Hearables}

\multauthor
{Mattes Ohlenbusch$^{1*}$ \hspace{1cm} Christian Rollwage$^1$ \hspace{1cm} Simon Doclo$^{1,2}$} { 
  $^1$ Fraunhofer Institute for Digital Media Technology IDMT,\\
Oldenburg Branch for Hearing, Speech and Audio Technology HSA, Germany\\
$^2$ University of Oldenburg, Department of Medical Physics and Acoustics\\
and Cluster of Excellence Hearing4all, Germany
\correspondingauthor{mattes.ohlenbusch@idmt.fraunhofer.de}{Mattes Ohlenbusch et al.}
}

\begin{document}

\maketitle

\begin{abstract}
Many hearables contain an in-ear microphone, which may be used to capture the own voice of its user in noisy environments.
Since the in-ear microphone mostly records body-conducted speech due to ear canal occlusion, it suffers
from band-limitation effects while only capturing a limited amount of external noise. To enhance the quality
of the in-ear microphone signal using algorithms aiming at joint bandwidth extension, equalization, and noise reduction,
it is desirable to have an accurate model of the own voice transfer characteristics between the entrance of the ear
canal and the in-ear microphone. Such a model can be used, e.g., to simulate a large amount of in-ear recordings to train supervised learning-based algorithms. 
Since previous research on ear canal occlusion suggests that own voice transfer characteristics depend on speech content, in this contribution we propose a speech-dependent system identification model based on phoneme recognition. 
We assess the accuracy of simulating own voice speech by speech-dependent and speech-independent modeling and investigate how well modeling approaches are able to generalize to different talkers.
Simulation results show that using the proposed speech-dependent model is preferable for simulating in-ear recordings compared to using a speech-independent model.
\end{abstract} 


\section{Introduction}\label{sec:introduction}
Hearables, i.e. smart earbuds containing a loudspeaker and one or more microphones, are often used in everyday noisy environments.
Although hearables are frequently used to enhance the voice of a person the hearable user is communicating with in a noisy environment, the scenario we are considering in this paper is to enhance the own voice of the user while talking in a noisy environment (e.g., to be transmitted via a wireless link to a mobile phone or another hearable).
In-ear microphones may offer benefits for own voice pickup since external noise is attenuated due to ear canal occlusion. However, own voice recorded inside the occluded ear suffers from amplification below 1\,kHz and heavy attenuation above 2\,kHz, leading to a limited bandwidth~\cite{bouserhal_-ear_2017}. 
The occlusion effect is affected by the ratio between the airborne and body-conducted component of own voice, which depends on the phonemes uttered by the user~\cite{reinfeldt_hearing_2010,saint-gaudens_towards_2022}.
Body conduction from different places of excitation and mouth movements during articulation likely influence this transmission behavior, which we refer to as transfer characteristics in this work. 
Additionally, body-produced noise (e.g., breathing sounds, heartbeats) may be recorded by an in-ear microphone~\cite{bouserhal_-ear_2019}. 
To enhance the quality of the in-ear microphone signal, several approaches have been proposed aiming at bandwidth extension, equalization and/or noise reduction, either based on signal processing~\cite{bouserhal_-ear_2017}
or supervised learning~\cite{ohlenbusch_training_2022}.
For supervised learning-based approaches large amounts of training data are typically required, which may be hard to obtain for realistic in-ear recordings.

Similar requirements have been addressed in supervised learning-based speech enhancement approaches using bone-conduction sensors. 
In~\cite{wang_multi-modal_2022}, it has been proposed to use a device-specific corpus of recordings to train a deep neural network (DNN) combining bone- and air-conducted speech recordings. In~\cite{wang_fusing_2022}, a semi-supervised training scheme has been utilized to jointly train a DNN simulating bone-conduction and a multi-modal enhancement DNN together. 
In~\cite{pucher_conversion_2021}, it has been proposed to convert airborne to bone-conducted speech using a DNN that accounts for individual differences between talkers using a speaker identification system. 
In previous work, we have proposed to estimate the transfer characteristics between the entrance of the ear canal and the in-ear microphone using a time-invariant linear model to simulate short segments of in-ear speech for data augmentation in DNN training~\cite{ohlenbusch_training_2022}.

In this paper, we propose to model own voice transfer characteristics using a phoneme-dependent system identification approach, where for each phoneme a different linear filter is estimated.
The proposed approach can be utilized to simulate speech at an in-ear microphone from regular speech recordings. 
%

\section{Signal Model}
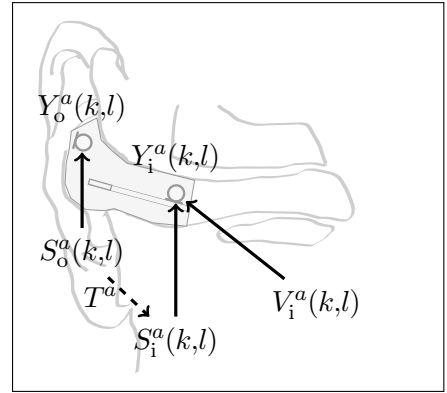
\begin{figure} 
 \centerline{\framebox{
    \centering 
    \begin{tikzpicture}
        \node[opacity=.4] (tepear) at (0,0) {\includegraphics[width=5cm]{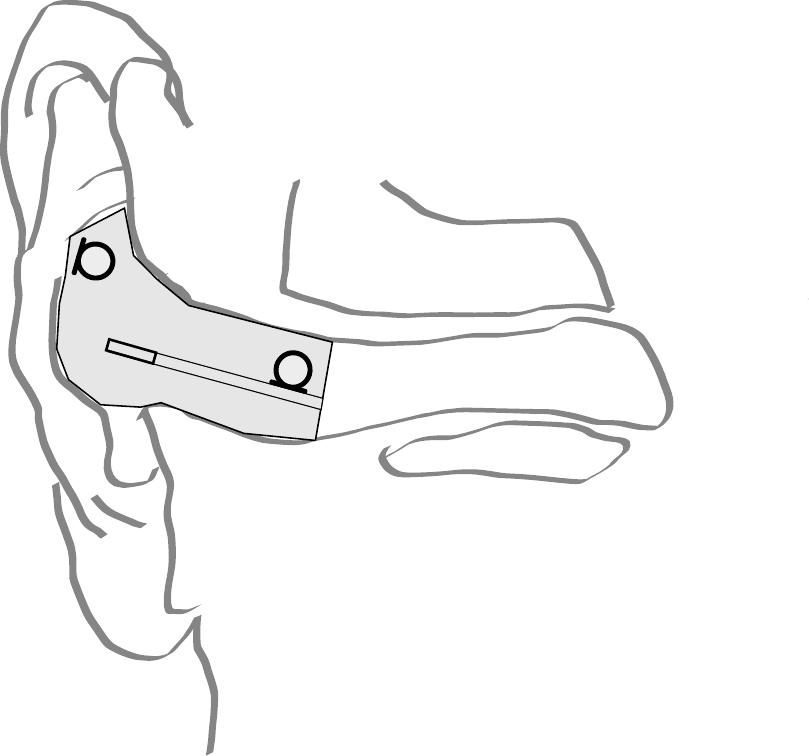}};
        \node[label={$Y_\mathrm{o}^a(k,l)$}] (oem) at (-1.925,0.725) { };
        \node[label={$Y_\mathrm{i}^a(k,l)$}] (iem) at (-0.6925,0.05) { };
        
        \node[below=of oem] (speech) {$S_\mathrm{o}^a(k,l)$};
        \node[below=1.5 of iem] (iemspeech) {$S_\mathrm{i}^a(k,l)$};
        \node[below right= of iem] (bodnoise) {$V_\mathrm{i}^a(k,l)$};

        \draw[->,very thick] (speech) -- (oem);
        \draw[->,very thick] (iemspeech) -- (iem);
        \draw[->, very thick] (bodnoise) -- (iem);
        \draw[->, dashed, very thick] (speech) -- (iemspeech) node[midway,left] {$T^a$};
        \vspace{-2cm}
    \end{tikzpicture}
    }}
    \caption{The signal model of own voice transfer characteristics considered in this paper.}
    \label{fig:sigmodel}
\end{figure}

Fig.~\ref{fig:sigmodel} depicts the considered scenario, where a talker is wearing a hearable device equipped with an in-ear microphone and an outer microphone, denoted by subscript $\mathrm{i}$ and $\mathrm{o}$, respectively. 
In the short time Fourier transform (STFT) domain the recorded own voice signal at the outer microphone of talker $a$ is denoted by $Y_\mathrm{o}^a(k,l)$ with $k$ the frequency bin index and $l$ the time frame index. 
The signal recorded at the in-ear microphone $Y_\mathrm{i}^a$ is assumed to contain an in-ear own voice speech component $S_\mathrm{i}^a$ and a body-noise component $V_\mathrm{i}^a$, i.e. 
\begin{equation}
    Y_\mathrm{i}^a(k,l) = S_\mathrm{i}^a(k,l) + V_\mathrm{i}^a(k,l)
\end{equation}
where $S_\mathrm{i}^a$ and $V_\mathrm{i}^a$ are assumed to be uncorrelated.
The own voice speech at the in-ear microphone $S_\mathrm{i}^a(k,l)$ is related to the own voice speech at the outer microphone $Y_\mathrm{o}^a(k,l)$ by transfer characteristics $T^a\{\cdot\}$, i.e.
\begin{equation}
    S_\mathrm{i}^a(k,l) = T^a\left\{Y_\mathrm{o}^a(k,l)\right\}.
\end{equation}
We assume these transfer characteristics to be linear, time-varying and individual.
The goal in this paper is to obtain an accurate transfer characteristics model, which is robust against talker mismatch. 

\section{Modeling Own Voice Transfer Characteristics}
\begin{figure} 
 \centerline{\framebox{
 \includegraphics
 [width=7.8cm]
 {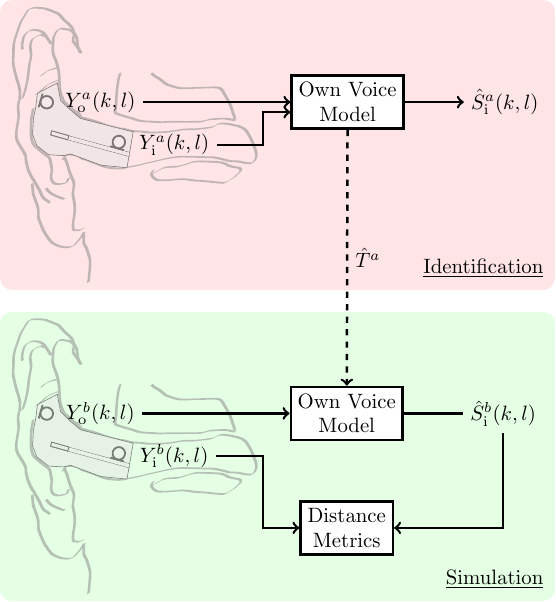}}}
 \caption{Overview of the identification and simulation steps for modeling own voice transfer characteristics.}
 \label{fig:simulation_diagram}
\end{figure}

In this section, we present two system identification-based approaches to model the transfer characteristics (see Fig.~\ref{fig:simulation_diagram}).
During identification, recordings of talker $a$ are used to obtain the model $\hat{T}^a$. 
During simulation, this model is used to obtain an estimate $\hat{S}_\mathrm{i}^b(k,l)$ of the own voice of talker $b$. If $a=b$, the model is applied to the same talker as in identification. If $a\neq b$, the model is applied to a different talker, i.e. talker mismatch is present. 
In Section~\ref{sec:speechindepmodel}, we present a time-invariant speech-independent model; in Section~\ref{sec:speechdepmodel} we propose a time-varying speech-dependent model.

\subsection{Speech-independent model}
\label{sec:speechindepmodel}
Assuming a linear time-invariant filter, the transfer characteristics $T^a$ of the speech-independent model are modeled as a relative transfer function (RTF) between the outer microphone and the in-ear microphone.
Since the outer microphone signal does not contain any additive noise, the RTF $\hat{H}^a(k)$ can be estimated using the well-known least squares approach~\cite{avargel_multiplicative_2007}, i.e.
\begin{equation}
\hat{H}^a(k) = \frac{ \sum_l Y_\mathrm{o}^{a,*}(k,l) \cdot Y_\mathrm{i}^a(k,l) }{ \sum_l |Y_\mathrm{o}^a(k,l)|^2}
\end{equation}
where $\cdot^*$ denotes complex conjugation.
For simulation, own voice speech of talker $b$ recorded at the outer microphone is filtered in the STFT domain as
    \begin{equation}
        \hat{S}_\mathrm{i}^b(k,l) = \hat{H}^a(k) \cdot Y_\mathrm{o}^b(k,l).
    \end{equation}

\subsection{Speech-dependent model}
\label{sec:speechdepmodel}
Since own voice transfer characteristics likely depend on speech content, we propose a time-varying speech-dependent model for the transfer characteristics $T^a$.
Using a phoneme recognition system,
we first obtain a frame-wise phoneme annotation $p(l) \in 1,\dots,P$ of a recorded speech signal, with $P$ possible phoneme classes.
For each unique phoneme $p^\prime$, an RTF is then estimated over all detected occurrences of this phoneme within the identification utterances of talker $a$, i.e. 
\begin{equation}
    \hat{H}_{p^\prime}(k) = \frac
    {\sum_{p(l)=p^\prime}  Y_\mathrm{o}^{a,*}(k,l) \cdot Y_\mathrm{i}^a(k,l)}
    {\sum_{p(l)=p^\prime} |Y_\mathrm{o}^a(k,l)|^2}.
\end{equation}
In total, the speech-dependent model hence consists of a database of $P$ RTFs.

For simulation, the phoneme sequence $p^b(l)$ is first determined on the own voice speech of talker $b$ recorded at the outer microphone. For each frame, the corresponding RTF $\hat{H}_{p(l)}^a(k)$ is then used to filter this signal in the STFT domain, i.e.
\begin{equation}
    \hat{S}_\mathrm{i}^b(k,l) =  \hat{H}_{p^b(l)}^a(k) \cdot Y_\mathrm{o}^b(k,l).
    \label{eq:simulation_speechdep}
\end{equation}

\section{Technical Evaluation}
In this section, the previously described transfer characteristics models are evaluated in terms of their accuracy in predicting own voice signals at an in-ear microphone.

\subsection{Evaluation data and setup}
A dataset of own voice speech from 14 native German talkers with approximately 30 minutes of speech in total is utilized in the evaluation. The hearable device used for recording is the closed-vent variant of the Hearpiece~\cite{denk_one-size-fits-all_2019}. Approximately 23 utterances per talker were recorded, resulting in a total of 329 utterances.
For the speech-dependent model, an in-house proprietary phoneme recognition system with $P=62$ phoneme classes was utilized, which was trained on German speech. 
Both the speech-independent and the speech-dependent model were estimated on all utterances per talker. For simulation, in-ear speech was predicted for each utterance of 5\,s length. No voice activity detection was employed so that utterances may contain small pauses between words.
The estimation accuracy is computed in two conditions: estimating speech of the same talker ($a=b$) for each utterance of each talker,
and estimating speech with talker mismatch ($a \neq b$) each utterance of each talker is simulated using a randomly assigned different talker model.

The simulations were carried out at a sampling frequency of $5$\,kHz to account for the reduced bandwidth of the in-ear speech.
An STFT framework with a frame length of $K=256$ (corresponding to 51.2\,ms), a frame overlap of 50\,\% and a square-root Hann window for analysis and synthesis was used. 
Since body-conducted speech travels faster than airborne speech, a prediction delay of 11 samples was applied to the in-ear signals prior to identification and simulation to enable causal modeling. 
As evaluation metric, we utilize the log-spectral distance (LSD)~\cite{gray_distance_1976} between the real and simulated in-ear recordings, where a lower value indicates a more accurate estimate. 

\subsection{Results and discussion}

\begin{figure}
\centerline{\framebox{
\includegraphics[width=7.8cm]{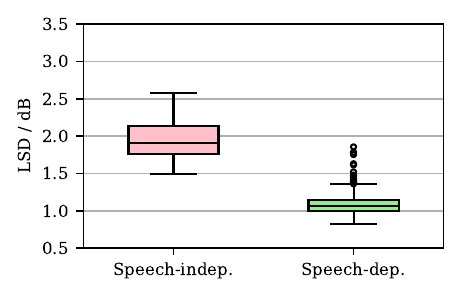}
}} 
\caption{LSD results for the same talker-condition.}
 \label{fig:lsd_ST_SU}
\end{figure}

The results for the same talker-condition are shown in Fig.~\ref{fig:lsd_ST_SU}. It can be observed that the in-ear speech signals can be predicted much better using the time-varying speech-dependent model than using the time-invariant speech-independent model.
Since in-ear recordings are assumed to also contain body noise uncorrelated to the outer microphone signals, remaining modeling errors are expected due to models not accounting for body-noise. 

\begin{figure}
 \centerline{\framebox{
 \includegraphics[width=7.8cm]{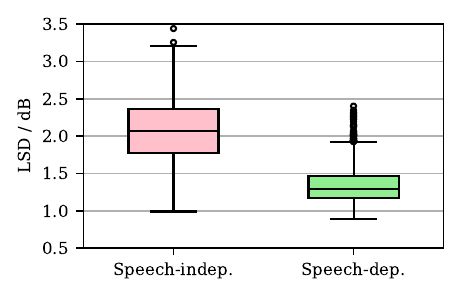}}}
 \caption{LSD results for the talker mismatch-condition.}
 \label{fig:lsd_DT_DU}
\end{figure}
The results for the talker mismatch-condition are shown in Fig.~\ref{fig:lsd_DT_DU}. 
It can be observed for both models that the performance is worse than in the same talker condition, where especially for the speaker-independent model there is a larger spread of LSD scores in the talker mismatch condition than in the same-talker condition. Nevertheless, also for the talker mismatch condition the speech-dependent model clearly outperforms the speech-independent model.

\section{Conclusion}
In this paper, two approaches to model own voice transfer characteristics in hearables have been investigated. Results indicate that speech-dependent modeling is beneficial compared to speech-independent modeling.

\section{Acknowledgments}
The Oldenburg Branch for Hearing, Speech and Audio Technology HSA is
funded in the program \frqq Vorab\flqq~by the Lower Saxony Ministry of Science and
Culture (MWK) and the Volkswagen Foundation for its further development. Part of
this work was funded by the German Ministry of Science and Education BMBF FK
16SV8811. This work was partly funded by the Deutsche Forschungsgemeinschaft
(DFG, German Research Foundation) - Project ID 352015383 (SFB 1330 C1).

\bibliography{zoterobib}

\end{document}